\documentclass[pdflatex,sn-mathphys]{sn-jnl}

\jyear{2022}%

\theoremstyle{thmstyleone}%
\theoremstyle{thmstyletwo}%

\theoremstyle{thmstylethree}%

\title[Turbulent Dissipation Through Cyclotron Resonance]{Mediation of Collisionless Turbulent Dissipation Through Cyclotron Resonance}
\begin{document}


\author*[1]{\fnm{Trevor A.} \sur{Bowen}}\email{tbowen@berkeley.edu}
\author[1,2]{\fnm{Stuart D.} \sur{Bale}}
\author[3]{\fnm{Benjamin D.G.} \sur{Chandran}}
\author[4]{\fnm{Alexandros} \sur{Chasapis}}
\author[5]{\fnm{Christopher H.K.} \sur{Chen}}
\author[6]{\fnm{Thierry} \sur{Dudok de Wit}}
\author[1]{\fnm{Alfred } \sur{Mallet}}
\author[7]{\fnm{Romain} \sur{Meyrand}}
\author[7]{\fnm{Jonathan } \sur{Squire}}

\affil*[1]{\orgdiv{Space Sciences Laboratory}, \orgname{Univeristy of California, Berkeley}, \orgaddress{\street{7 Gauss Way}, \city{Berkeley}, \postcode{94720}, \state{CA}, \country{USA}}}
\affil[2]{\orgdiv{Physics Department}, \orgname{Univeristy of California, Berkeley}, \orgaddress{\city{Berkeley}, \postcode{94720}, \state{CA}, \country{USA}}}
\affil[3]{\orgdiv{Department of Physics \& Astronomy}, \orgname{University of New Hampshire}, \orgaddress{\city{Durham}, \postcode{03824}, \state{NH}, \country{USA}}}
\affil[4]{\orgdiv{Laboratory for Atmospheric and Space Physics}, \orgname{University of Colorado}, \orgaddress{\city{Boulder}, \postcode{80303}, \state{CO}, \country{USA}}}
\affil[5]{\orgdiv{Department of Physics and Astronomy, London E1 4NS, UK}, \orgname{ Queen Mary University of London}, \orgaddress{ \city{London}, \postcode{E1 4NS}, \country{UK}}}
\affil[6]{\orgname{LPC2E, CNRS and University of Orl\'{e}ans}, \city{Orl\'{e}ans}, \postcode{610101}, \country{France}}
\affil[7]{\orgdiv{Physics Department}, \orgname{University of Otago}, \orgaddress{\city{Dunedin}, \postcode{9010}, \country{New Zeland}}}

\abstract{The dissipation of magnetized turbulence is fundamental to understanding energy transfer and heating in astrophysical systems ranging from the solar wind and corona to accretion disks and the intracluster medium. Collisionless interactions, such as resonant wave-particle process, are known to play a role in shaping these turbulent astrophysical environments. While evidence suggests that ion-cyclotron resonant interactions contribute to collisionless plasma heating in the solar wind and corona, a direct coupling between magnetized turbulence and ion cyclotron waves has yet to be demonstrated. Here, we present evidence for the mediation of turbulent dissipation in the solar wind by ion-cyclotron waves. Our results show that ion-cyclotron waves interact strongly with magnetized turbulence, indicating that they serve as a major pathway for the dissipation of large-scale electromagnetic fluctuations. We further show that the presence of cyclotron waves significantly weakens observed signatures of intermittency in sub-ion-kinetic turbulence, which are known to be another pathway for dissipation. These observations results suggest that in the absence of cyclotron resonant waves, non-Gaussian, coherent structures are able to form at sub-ion-kinetic scales, and are likely responsible for turbulent heating. We further find that the cross helicity, i.e. the level of Alfv\'enicity of the fluctuations, correlates strongly with the presence of ion-scale waves, demonstrating that dissipation of collisionless plasma turbulence is not a universal process, but that the pathways to heating and dissipation at small scales are controlled by the properties of the large-scale turbulent fluctuations. We argue that these observations support the existence of a helicity barrier, in which highly Alfv\'enic, imbalanced, turbulence is prevented from cascading to sub-ion scales thus resulting in significant ion-cyclotron resonant heating. Our results may serve as a significant step in constraining the nature of turbulent heating in a wide variety of astrophysical systems.}

\maketitle

\paragraph{Introduction} 
Turbulence is an important means of energy transfer in astrophysical environments \citep{Higdon1984,Quataert1998,Cranmer2000,CranmervanBallegooijen2003,Zhuravleva2014,Chen2019}. In hydrodynamics, turbulence manifests as the nonlinear shearing of structure in the velocity field (i.e. eddies) into smaller scales, which continues until structures become small enough for viscosity, mediated by inter-particle collisions, to efficiently convert energy stored in the eddies into thermal energy, dissipating the turbulence \cite{Frisch1995}.  
Hydrodynamic turbulence is governed by the energy cascade rate, i.e. the dissipation of large scale eddies, as well as the fluid-viscosity at microphysical scales. The independence of the energy cascade rate from the viscosity, which is a result of continuous generation of continuously smaller scales via nonlinearities, gives hydrodynamics a universal nature.
 In contrast, astrophysical plasmas are often collisionless and thus require more exotic processes to dissipate turbulent energy into thermal energy of the constituent particles.


Our understanding of collisionless turbulent dissipation, and its importance in heating astrophysical systems \cite{Coleman1968,Belcher1971, Cranmer2000,CranmervanBallegooijen2003, Higdon1984,Quataert1998,Zhuravleva2014,Chen2019}, is constrained observationally almost entirely by in situ measurements of terrestrial and heliospheric plasma environments. Spacecraft observations of such plasmas have proven significant test-bed for diagnosing the relevant collisionless process resulting in particle heating and acceleration, improving our understanding of collisionless heating and turbulent dissipation both within our solar system and in far-off astrophysical environments \cite{Higdon1984,Quataert1998,Zhuravleva2014}. 

The heating and acceleration of the solar wind and the solar corona through collisionless mechanisms remains a poorly understood, yet fundamental, process with analogous dynamics occurring in many astrophysical systems. As plasma leaves source regions on the solar surface it undergoes continuous heating, resulting in a hot and tenuous upper atmosphere known as the solar corona. At coronal temperatures, the plasma cannot be confined by the Sun's gravity and is accelerated into a super-sonic solar wind that streams into the solar system,  eventually reaching the interstellar medium \cite{Parker1958}. This process largely occurs without inter-particle collisions, with the source of energy responsible for heating and acceleration of the solar wind and corona clearly lying within the sun's magnetic field. Understanding these processes is primary goal of NASA's Parker Solar Probe (PSP) mission \cite{Fox2016}.  


Electromagnetic interactions in plasmas sustain a variety of modes, i.e. waves, that provide a number of ways for the energy stored in electromagnetic fields and plasma flows at large scales to dissipate into thermal energy. At scales much larger than the ion gyroradius $\rho_i = v_{thi}/\Omega_i$, where the ion thermal speed is $v_{thi}=\sqrt{2T_{i}/m_i}$ and the ion gyrofrequency is $\Omega_i=ZeB_0/m_i$, magnetized turbulence is often polarized perpendicular to the mean background magnetic field $\mathbf{B}_0$, and approximately satisfies $\delta \mathbf{v_\perp} \approx \pm \delta \mathbf{b_\perp}$ \citep{Belcher1971} with $\mathbf{\delta b_\perp}=\mathbf{\delta B_\perp}/\sqrt{\rho_0\mu_0}$, where $\rho_0$ is the average plasma mass-density. These qualities, common to Alfv\'{e}n waves \cite{Alfven1942}, lead to such turbulence being described as ``Alfv\'{e}nic'' in nature. The strength of the ominant Alfvén mode relative to subdominant fluctuations is referred to as the ``imbalance'' and can be quantified using the normalized cross helicity \begin{align} \sigma_c =\frac{2\langle\delta \mathbf{v_\perp} \cdot \delta \mathbf{b_\perp}\rangle}{\langle\delta \mathbf{v_\perp}^2\rangle + \langle\delta \mathbf{b_\perp}^2\rangle}, \end{align} where values of $\sigma_c\pm 1$ indicate highly Alfv\'{e}nic fluctuations.

Turbulent dissipation at large scales, which are often approximated using magnetohydrodynamics, is negligible. Thus, in order to dissipate, energy must be transferred to small-scale fluctuations that are comparable to $\rho_i$ and the ion-inertial length $d_i=\rho_i/\sqrt{\beta_i}$, where $\beta_i=2n_0\mu_0T_i/B_0^2$. Near these ion-kinetic scales, Alfv\'{e}n waves become dispersive \cite{Lysak1996} and can be damped, heating the plasma, while simultaneously transferring energy into a kinetic-Alfv\'{e}n-wave (KAW) cascade \cite{Bale2005,Schekochihin2009,Salem2012,Chen2013}. 
In addition to kinetic-Alfv\'{e}n-wave turbulence, plasmas sustain a variety of other electromagnetic waves \cite{Gary1993} that may play a role in plasma heating through wave-particle interactions\cite{Barnes1966,Denskat1983,Goldstein1994,Howes2008,Chandran2010a,Chandran2010b,Chandran2013,Leamon1998a, Leamon1998b,parashar2015,Mallet2017,Chen2019}. 


Numerous populations of circularly-polarized, electromagnetic waves have been documented in the solar wind at ion-kinetic scales \cite{Jian2009,Bowen2020a}. These waves, which interact efficiently with particles via instabilities \citep{Gary1993} and directly heat plasma through resonant coupling to particle gyromotion, are one potential pathway to turbulent dissipation \cite{Denskat1983,Leamon1998a,Bowen2022}. 
The observed temperature anisotropy of ions in the solar corona is consistent with cyclotron resonant heating \citep{Kohl1997}, and a number of models invoke cyclotron heating as a means of dissipating turbulent energy into coronal-ion populations \cite{Cranmer2000,CranmervanBallegooijen2003}. While there is no direct evidence for coronal heating through cyclotron resonance, substantial evidence exists for active cyclotron resonance in the solar wind \cite{MarschTu2001,Smith2012,He2015,Bowen2022}. 

Here, we demonstrate the connection between the presence of parallel-propagating left-hand-polarized waves, a proxy for ion-cyclotron waves (ICWs) \cite{Bowen2020d}, and signatures of turbulent dissipation. We show that signatures of turbulent dissipation are mediated by the presence of ICWs, providing direct evidence for the role of cyclotron resonance in turbulent dissipation. In the absence of ICWs, we find that the range of turbulence  much smaller than the ion-kinetic scales is populated with non-Gaussian, intermittent, fluctuations, indicative of small-scale current sheets  \cite{Sorriso-Valvo1999, parashar2015,Chhiber2021} and KAW turbulence \cite{Bale2005,Schekochihin2009,Salem2012} that are likely responsible for dissipation. These results suggest that turbulent dissipation in magnetized plasmas is not a universal process \cite{Frisch1995}, and that the nature of dissipation depends on large scale configuration of the plasma. Our observations are consistent with a proposed ``helicity-barrier'' mechanism \cite{Meyrand2021}, in which conservation of energy and helicity prevent highly Alfv\'{e}nic turbulence from cascading to sub-ion scales, resulting in the generation of ICWs that heat low-$\beta$ magnetized plasmas \cite{Squire2022}.

\section{Results}

Fig. 1 shows in situ PSP observations for the full day of 2021-09-27 \cite{Bale2016,Kasper2016}. Data are shown in the Radial-Tangential-Normal (RTN) coordinate system.  The spacecraft was located $\sim 30 R_\odot$ from the sun. Fig. 1(a-e) show the measured solar wind velocity $\mathbf{V}_{sw}$, magnetic field $\mathbf{B}$, scalar proton temperature $T_i$, electron density $n_e$, $\beta_i$, and the angle between the magnetic field and solar wind flow direction $\lvert \theta_{BV}\rvert$. We measure a mean speed of $\langle V_{sw}\rangle$=355 km/s; $\langle T_i \rangle$=55 eV; average magnetic field magnitude of $\langle\lvert B\rvert\rangle=255$ nT; $\langle n_e\rangle$ of 1200 cm$^{-3}$; and an average $\langle\beta_i\rangle=0.43$. 

\begin{figure}[h]%
\centering
\includegraphics[width=0.9\textwidth]{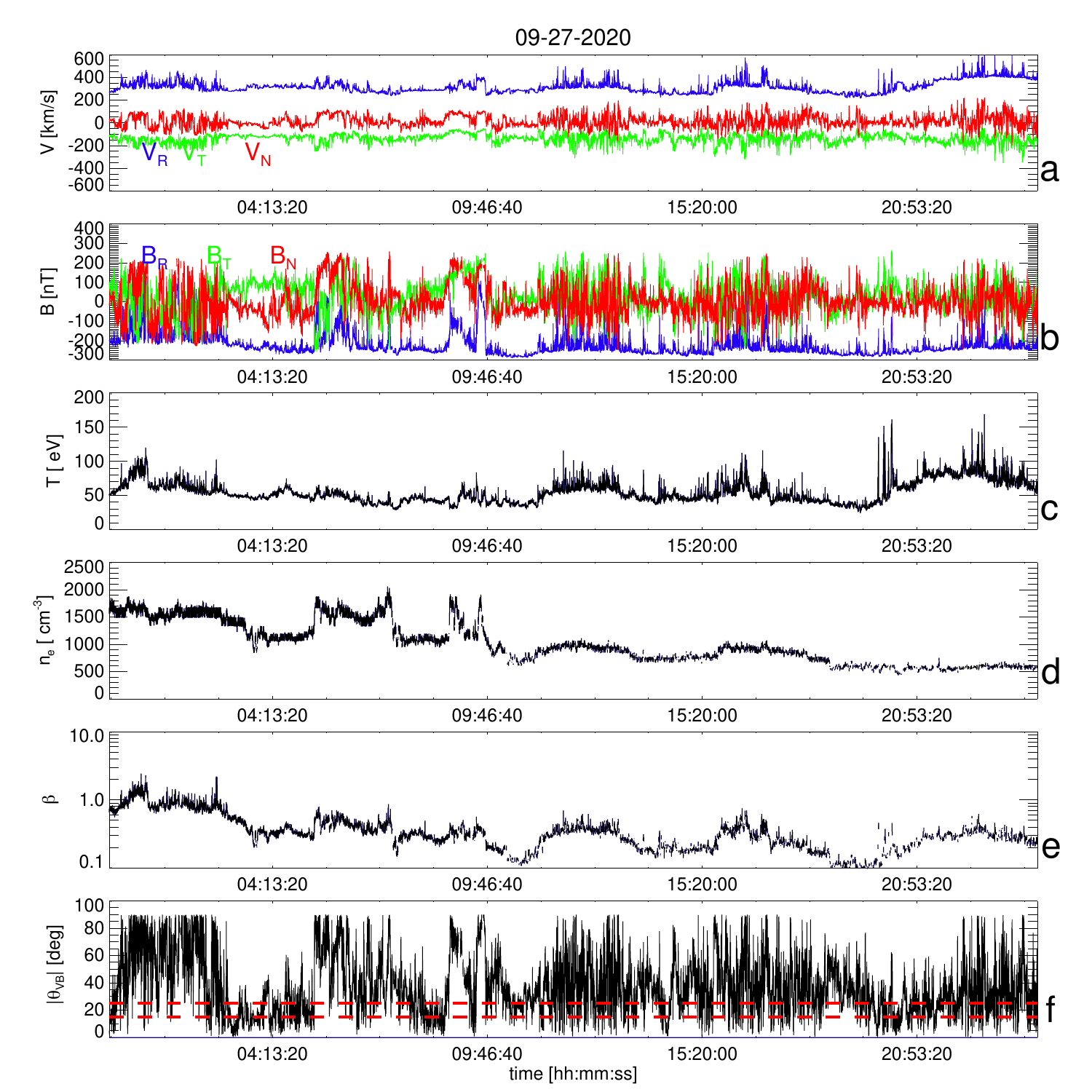}
\caption{Survey plot of PSP FIELDS and SWEAP data on 2021-09-27. Panels show a) SPANi proton velocity data in RTN coordinates; b) FIELDS magnetometer measurements in RTN coordinates; c) proton temperature; d) FIELDS RFS QTN total electron density $n_e$; e) proton $\beta$; f) angle between $V_{sw}$ and $B$. Red dashed lines in panel (f) show range of $\theta_{BV}$ used to select the intervals studied.\label{fig1}}
\end{figure}

\begin{figure}[h]%
\centering
\includegraphics[width=0.9\textwidth]{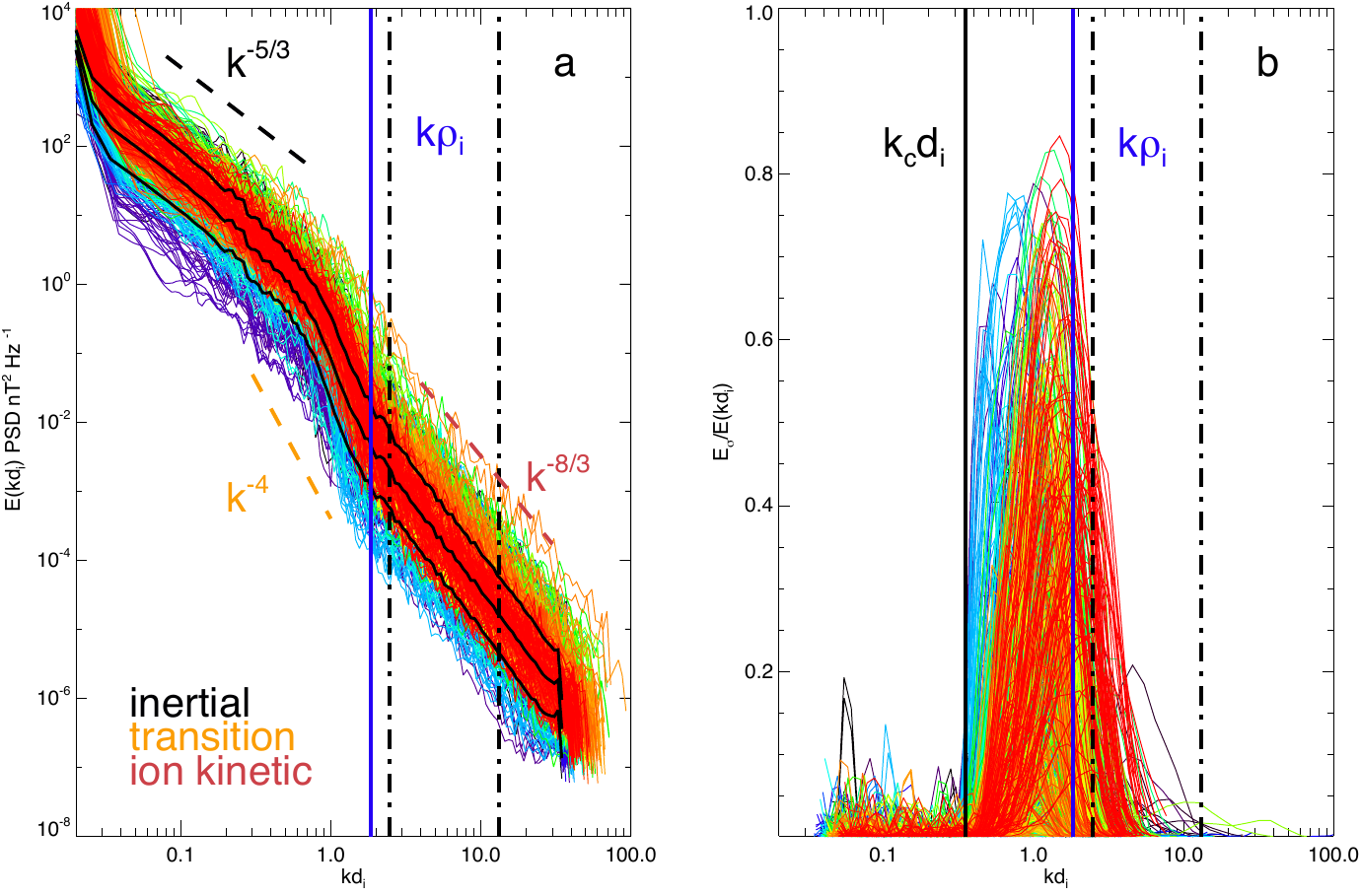}
\caption{a) Observed trace magnetic field spectra for 747 intervals. Spectra are computed from measured time series using FFT methods, frequencies are then normalized to the ion inertial scale $kd_i$ using the Taylor hypothesis. Dashed lines show various known spectral scalings: $k^{-5/3}, k^{-8/3},k^{-4}$. b) Circularly polarized power normalized to trace power spectra computed from a wavelet transform; frequencies are normalized to $kd_i$ using the Taylor hypothesis. \label{fig2}. The solid vertical line shows an approximate lower bound cutoff of the circularly polarized power $k_cd_i=0.35 $. The blue line shows the average position of the ion-gyroscale  $k\rho_i$ with respect to the ion inertial scale $kd_i$.}
\end{figure}

We study turbulent properties in this stream using a set of 747 power-spectral densities of trace magnetic field fluctuations, $E(kd_i)$, from 2020-09-27, shown in Fig. 2(a). The average ion-inertial length 7.7 km with average solar wind speed of 347 km/s corresponds to a spacecraft frequency of 7.4 Hz with a standard deviation of 1.5 Hz. The average spectra and one standard deviation levels are shown as solid black lines. In the inertial range, the fluctuations are known to have an approximate power-law scaling with $k^{-\alpha}$; in agreement with previous results we find $\alpha$ is 3/2 or 5/3 for $k_\perp$ spectra, which is measured when  $\theta_{BV}$ is oblique, and $\alpha$ of 2 for $k_\parallel$ spectra, measured when $\theta_{BV}$ corresponds to parallel/anti-parallel angle \cite{GS95,Horbury2008}. There is often a transition range at ion-kinetic scales with a very steep $\alpha\approx 4$ scaling \cite{Sahraoui2009,Kiyani2009,Bowen2020c}, while at sub-ion-kinetic scales, a scaling with an approximate $\alpha\sim{-8/3}$ is observed \cite{Goldstein1994,Alexandrova2008,Sahraoui2009,BoldyrevPerez2012}.  
Corresponding power-law scalings of -5/3, -4 and -8/3 are plotted on Fig. 2(a) to highlight these regions.
Circular polarization is quantified using a Morlet wavelet transform (see Methods). Fig. 2(b) shows the circularly polarized power computed from a wavelet analysis for each of the 747 intervals. The power is normalized to the total wavelet power, i.e. a value of unity means that the power is entirely circularly polarized. 
We define a cutoff at $k_cd_i=0.35$, which is uniformly below the measured circularly polarized power in each interval. The blue line shows the average position of $k\rho_i$ with respect to $kd_i$.

Fig. 3(a) shows spectra of 150 intervals that satisfy  $15^\circ<\theta_{BV}<25^\circ$, which is imposed to control for effects of anisotropy. This angle range was chosen to enable observation of ICWs \cite{Bowen2020a}, but retaining some portions of the perpendicular spectra, which dominates the cascade \cite{Horbury2008} Fig. 3(a) shows power spectra with frequency normalized to $kd_i$, and power normalized to $E(k_cd_i)$.

The ratio in power over the transition range, $\tilde{E}_{TK}=\text{min}[E(kd_i)]/E(k_cd_i)$,  is measured to quantify the drop in energy between transition and kinetic scales. Figure 3 is color-coded by $\tilde{E}_{TK}$, such that darker spectra have the greatest drop in power over the transition range, while lightest colors correspond to the smallest drop in power. We similarly measure  $\tilde{E}_{IK}=\text{min}[E(kd_i)]/E(0.1k_cd_i)$, as the drop in energy spectra between inertial and kinetic scales. Figure 3(b) shows the moving window estimate of the spectral index $\alpha$ as a function of $kd_i$. A clear steepening is observed near $kd_i\sim k_cd_i$, with a shallow spectra at $kd_i<k_cd_i$ and the recovery of a sub-ion kinetic range scaling $\alpha\sim 8/3$ to 3 at $kd_i\gg k_cd_i$ \cite{BoldyrevPerez2012}. The lowest frequency range has an average spectral index of around $-2$, consistent with known scalings of the spectra at $15^\circ$ \cite{Horbury2008}, although we note that there is significant variation in the low frequency spectral scaling. 

The fraction of circularly polarized power, $E_\sigma/E_\perp$, in each interval is determined using a Morlet wavelet transform (\cite{TorrenceCompo1998,Bowen2020a},  see Methods). Figure 3(c) and Fig. 3(d) show the normalized $E_\sigma/E_\perp$ from the wavelet transform for the left- and right-handed polarizations respectively. Very little right-handed power is observed. The color coding shows that intervals with greater drops in power, a proxy for turbulent dissipation \cite{Bowen2020c}, coincide with significant left-hand-circularly-polarized power. We compute several correlations using the nonparametric Spearman-ranked correlation coefficient, $R$, highlighting the connection between circular polarization and turbulent properties. 

Figure 4(a) shows ratios of the energy spectra $\tilde{E}_{TK}$ and $\tilde{E}_{IK}$ plotted against the maximum circular polarization. Strong correlations between the drop in power and presence of left handed waves are present both for the drop over the transition and kinetic ranges, i.e. $\tilde{E}_{TK}$. ($R=-0.6$), and for the difference in energy between the inertial and the kinetic scales, $\tilde{E}_{IK}$ ($R=-0.8$). Figure 4(b) shows that $\tilde{E}_{TK}$ is significantly correlated ($R=-0.7$), to the level of cross helicity at inertial scales. Fig. 4(c) shows the correlation between the maximum circularly polarized power $\text{max}[E_\sigma/E_\perp]$ and the cross helicity, $\sigma_c$. The strong correlation ($R=0.6$), indicates that circularly polarized power is most often found in higher cross helicity intervals. Figure 4(d) shows the correlation between the kinetic-range spectral index, computed between 3.5$kd_i$ and 12$kd_i$, and the cross helicity. Higher cross helicity is moderately correlated with flatter kinetic-scale spectral indices (R=0.5). Together, the panels in Fig. 4 show that the turbulent dynamics at sub-ion scales are related to the presence of left-handed waves at ion scales as well as to the MHD-scale cross helicity, which is a measurement of the imbalance of the turbulence. Furthermore, we show that the left-handed waves are highly correlated with high cross helicity states, indicating that these intervals are preferentially associated with cyclotron-waves that can serve as a pathway to dissipation \cite{Bowen2022}. While the existence of ICWs may not inevitably result in the irreversible thermalization of turbulent energy, i.e. heat, we further study their connection to dissipation signatures through studying their association with intermittent scaling of the kinetic scale turbulence. 

While Fig. 4 highlights the connection between ion-scale waves and the spectral energy density, further evidence for the impact of the waves on kinetic scale, i.e. $k\rho_i >1$, turbulent fluctuations exists in their intermittent scaling. Figure 5 shows quantities relating to the kinetic-scale cascade. The distribution of increments $P({\Delta B}_i^\tau)$ at $\tau= 0.01s$  (see Methods) for a single direction of the magnetic field measurements are shown in Fig. 5(a). The colors again correspond to the drop in energy spectra over transition to kinetic scakes, $\tilde{E}_{TK}$, with darker colors associated with larger drops in power. Each $P({\Delta B}_i^\tau)$ is normalized to the standard deviation of the distribution.  There is a clear relationship between $P({\Delta B}_i^\tau)$ and $\tilde{E}_{TK}$, which are highly correlated with ion-scale waves. Specifically, lighter colors, i.e. smaller drops in power, have more significant non-Gaussian tails.  This is further evidenced in Fig. 5(b), which shows that the kurtosis for the kinetic-scale magnetic field measurements is correlated with  $\text{min}[E(kd_i)]/E(k_cd_i)$: larger drops in the energy spectra, which Fig. 3 shows are associated with the presence of ICWs, have a lower kurtosis in the kinetic range.

Fig. 5(c) shows the structure function scaling exponent $\zeta(p)$ (see Methods) computed for each interval over in the kineti-scale range of $0.007<\tau<0.027$ s, frequencies from 18.3-97 Hz. This range is above the scales studied in our wavelet analysis (0.36-17.5 Hz) and thus does not overlap with ion-kinetic scales where circular polarization is present; accordingly our analysis not include intermittent signatures brought on by the occurrence of the waves themselves. Error bars are not included on the data in Fig. 5(c) for clarity, though estimated levels of mean and maximum errors in the scaling exponent are shown. The derivative $d{\zeta}/d{p}$ is shown in Fig. 5(d). Darker colors are observed to be more self-similar and less intermittent, whereas the lighter colors show more significant deviations from self-similar scaling, indicating intermittency of the fluctuations. 

The curvature of the scaling exponents $\Gamma_{\zeta}$ is computed for each interval as a  measure of the nonlinearity in $\zeta$, where larger $\Gamma_{\zeta}$ indicates less linear scaling in $\zeta(p)$, and thus more intermittency in the fluctuations. Fig. 5 (e-f) show the average  $\Gamma_{\zeta}$ from each interval plotted against $\tilde{E}_{TK}$, where a R=0.7 Spearman-ranked correlation is obtained, as well as $\text{max}[E_\sigma/E_\perp]$, where  R=-0.6. These significant correlations indicate that kinetic range intermittency is much stronger intervals without significant wave activity, with shallower drops in power over the transition range.

\begin{figure}[h]%
\centering
\includegraphics[width=0.9\textwidth]{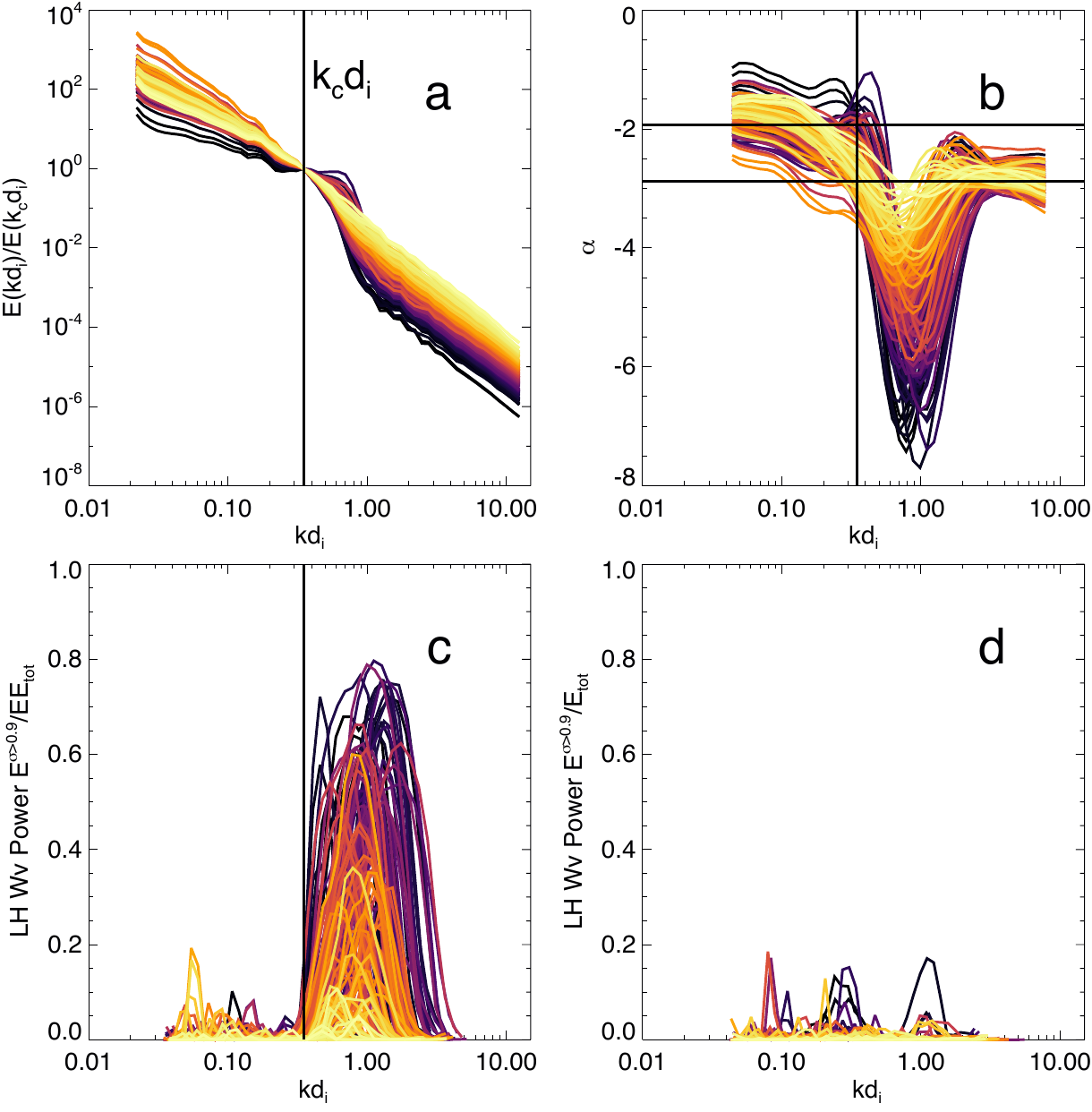}
\caption{a) Turbulent spectra from intervals with $15^\circ<\theta_{BV}<25^\circ$ normalized to unity power at $f_c$. Spectra are color coded by the difference in energy between $f_c$ and $35 f_c$, darker colors have a greater drop in power. b) Moving window spectral index for the 150 spectra, a break is observed near $0.35f_c$, which corresponds to the onset of circularly polarized power; recovery to a power law scaling of approximately -8/3 is recovered at sub-ion scales. Mean values of $\alpha$ in the inertial and kinetic ranges are shown as black horizontal lines. (c) Circularly polarized power in left handed modes normalized to total power. (d) Circularly polarized power in right handed modes normalized to total power. }\label{fig3}
\end{figure}

\begin{figure}[h]%
\centering
\includegraphics[width=0.7\textwidth]{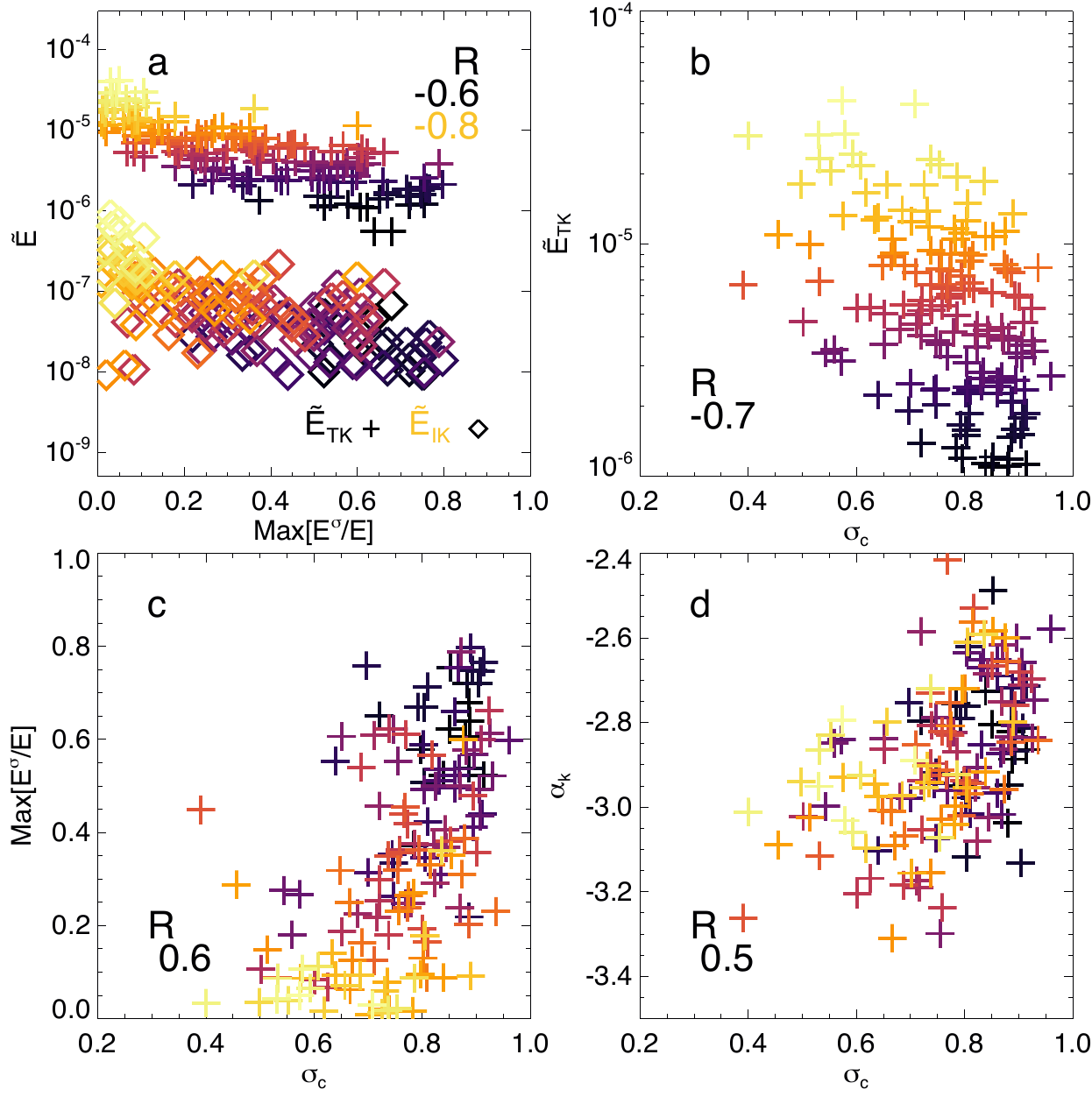}
\caption{a) Scatter plot of drop measured turbulent power against level of circular polarization; + shows ratio in power between $f_c$ and 35 $f_c$, corresponding to a drop in power between transition and sub-ion scales. Diamonds show the ratio between an inertial-range scale, $0.1 f_c$, and 35 $f_c$. Spearman correlations of -0.5 and -0.7 are measured. b) Drop in spectral power between $f_c$ as function of measured cross helicity. c) Level of circularly polarized power against cross helicity. d) Measured kinetic-scale spectral index, $\alpha_k$, (computed between 3.5$kd_i$ and 12$kd_i$) against cross helicity. Dashed lines show linear least square fit lines. Colors for each data point correspond to the individual spectra shown in Fig. 3. }\label{fig4}
\end{figure}

\begin{figure}[h]%
\centering
\includegraphics[width=0.8\textwidth]{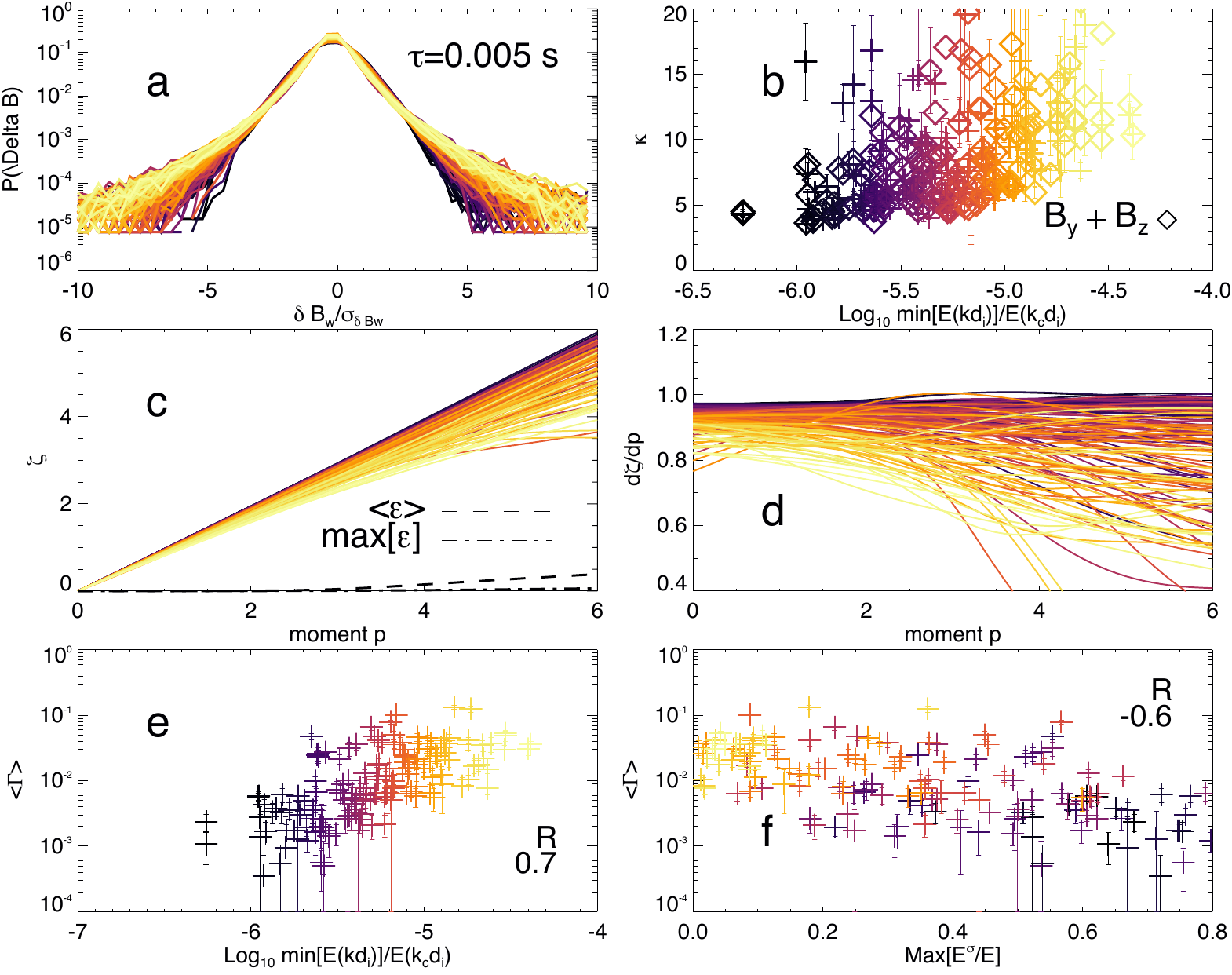}
\caption{a) Distributions of increments at $\tau=0.01$s for each interval in the SCM $\hat{z}$ direction. Each distribution is normalized to its standard deviation. Colors correspond to drop in energy $\tilde{E}_{TK}$ shown in Fig. 3. b) Kurtosis of the fluctuations in the SCM $\hat{y}$ and SCM $\hat{z}$ directions (+ and $\diamond$) at $\tau=0.01$s plotted against $\tilde{E}_{TK}$. Error bars show standard error on mean computed from bootstrapping subsets of data.  c) Intermittency of each interval is illustrated using scaling exponent of the structure functions $\zeta(p)$.   Average and maximum standard error for each moment p are shown as dashed and dot-dashed lines. d) Derivative $d\zeta/dp$, a linear $\zeta(p)$ (i.e. constant $d\zeta/dp$) corresponds to non-intermittent fluctuations. e) Average curvature of sub-ion-kinetic range scaling exponent $\Gamma_\zeta$ plotted against drop in spectra, $\tilde{E}_{TK}$.  f) Average curvature of sub-ion-kinetic range scaling exponent $\Gamma_\zeta$ plotted againstm aximum circularly polarized power at ion-kinetic scales.}\label{fig5}
\end{figure}
\section{Discussion}

These results show that left-handed waves interact strongly with turbulence at ion scales, with strong consequences for the sub-ion kinetic scale turbulence. Under the hypothesis that the solar wind consists of superposed non-interacting turbulent fluctuations and circularly-polarized waves, signatures of the turbulence deep into the sub-ion range should not correlate with the presence and properties of ion-scale waves. The strong correlation between the decrease in energy across the spectra with the level of left-handed circular polarization observed at ion-kinetic scales suggests that waves play an active role dissipating energy from the turbulent cascade at these scales. Whether this process happens through direct absorption from the turbulence via resonant interaction \cite{Denskat1983,Goldstein1994,Leamon1998a,Smith2012}, or the waves are generated via a secondary process following a primary mechanism (e.g. oblique cyclotron resonance or stochastic heating \cite{Chandran2010a,Chandran2010b,Squire2022}) the interaction between these waves and the turbulent cascade is well established by these observations.

Previous work has suggested that the variability in the transition-range spectral index may relate to properties of the inertial-range cascade \cite{Smith2006,Bruno2014}. Our results expand on these ideas providing direct evidence that the presence of left-hand polarized ion-scale waves at kinetic scales, which are likely responsible for significant ion heating \cite{Bowen2022}, is highly correlated with the inertial-range cross helicity. 

Furthermore, we show that the presence of the ion-scale waves is connected not just to the large scale fluctuations, but has significant correlation with signatures of sub-ion-scale turbulence. Analysis of structure function scaling exponents reveals that that intervals with significant wave activity have more self-similar, less intermittent, structure-function scalings. The sub-ion-kinetic scale kurtosis of these intervals is relatively low, suggesting that ion-scale dissipation may re-Gaussianize, i.e. randomize, the turbulent fluctuations \cite{Mallet2019,Chhiber2021}, which are known to be strongly intermittent in the inertial range \cite{Sorriso-Valvo1999}. Additionally, it seems that the presence of significant ion-scale wave activity inhibits the growth and production of intermittent structures, often associated with  current sheets \cite{Dudokdewit1996,Sorriso-Valvo1999,Kiyani2009,parashar2015,Chhiber2021,Roberts2022} deep into the kinetic cascade range. We speculate that the steep drops in power associated with significant ion-cyclotron waves near the transition range may cause the kinetic scale nonlinearities to weaken \cite{Howes2011,David2019}, such that nonlinear interactions no longer produce strongly intermittent fluctuations.


While the nature of kinetic-scale intermittency has been debated \citep{parashar2015,Mallet2019,Chhiber2021,Roberts2022}, 
and the exact means of dissipating small-scale current sheets remains largely unconstrained, it is clear that these coherent structures must play an active role in collisionless plasma heating \cite{Alexandrova2008,parashar2015,Boldyrev2012}. Simulations and observations suggest that such current sheets may dissipate their energy to electrons via Landau damping \cite{Howes2011,Boldyrev2012,TenBarge2013, Chen2019}. 
Our observation of the mediation of energy transfer via ion scale waves suggests that the fraction of turbulent energy deposited in the ions via cyclotron-resonant interactions versus that deposited in electrons via Landau damping almost certainly depends on the large-scale characteristics of the turbulence--i.e., particularly the cross helicity.


These results show that there is not a single, universal, description of the intermittent evolution of turbulence at the sub-ion scales, and that discrepant intermittent scalings observed at kinetic scales \cite{Kiyani2009,parashar2015,Chhiber2021,Roberts2022} relate to the nature of the turbulence at large scales and their available pathways to dissipation and heating at ion-scales. Previous work \cite{BoldyrevPerez2012} has suggested that intermittent effects may steepen kinetic-scale turbulence from a $k^{-7/3}$ to a $k^{-8/3}$ spectra. While our results show that intermittency may play a role in setting the kinetic-range spectral index, our measured scalings are characteristically steeper than the \cite{BoldyrevPerez2012} prediction of $k^{-8/3}$, which may be due to Landau damping of kinetic scale fluctuations \cite{Howes2011} or the weakening of the cascade \cite{Galtier2003,Meyrand2013,David2019}.


Our results support recent suggestions that turbulence in high-cross-helicity states may not be able to physically transfer the majority of turbulent energy to kinetic scales due to the need for simultaneous conservation of both helicity and energy \cite{Meyrand2021,Squire2022}. This mechanism, termed the helicity barrier\cite{Meyrand2021}, hypothesizes a build up of energy in the inertial range that results in the growth of fluctuations with small parallel scales, that damps the turbulence through cyclotron resonance \cite{IsenbergLee1996}. The helicity barrier allows only a small fraction of the flux from large scales to pass into the sub-ion cascade. Regions with higher cross helicity let through a smaller energy fraction, implying a larger spectral drop through the transition range and thus smaller-amplitude sub-ion turbulence. Sufficiently small amplitude sub-ion turbulence may have nonlinear interactions that are subdominant to the linear kinetic-Alfv\'{e}n-wave physics, thus rendering the turbulence weak. This may explain the correlation of the turbulent intermittency with the large-scale cross helicity and significant spectral drop $\Delta E$ over the transition range.

Recent work \cite{Bowen2022} has highlighted that ICWs are quasilinearly damped in the solar wind \cite{IsenbergLee1996}, with heating-rates that are a substantial fraction of the turbulent energy-cascade rate. While the physical processes that generate ICWs have not been concretely established, it is clear that they significantly impact the observed kinetic plasma distributions \cite{MarschTu2001,He2015,Bowen2022}, and as we show here, the nature of turbulent dissipation and sub-ion turbulent energy transfer. The dissipation of turbulent fluctuations at ion-kinetic scales is likely key to heating in the solar wind and corona.  While ICWs have long been suspected to play a role in heating collisionless environments, our work provides strong evidence for the interaction of cyclotron waves with astrophysical turbulence, suggesting that cyclotron resonance is a significant pathway in turbulent dissipation and heating in the inner heliosphere and other $\beta <1$ plasma environments.

That dissipative mechanisms at small scales depend on the large-scale geometry of the turbulent fluctuations, specifically the level of Alfv\'{e}nicity, is a striking contrast to hydrodynamic turbulence, where the independence of the cascade from its viscous dissipation is typically considered to be a universal process \cite{Meyrand2021}. While there is a longstanding debate regarding the universality of dissipation in magnetized turbulence \cite{Smith2006,Alexandrova2008,Kiyani2009,DudokdeWit2022}, our results show that the relative importance of different dissipation mechanisms depends on the large scale properties of the turbulence. Our results expand significantly on previous work by suggesting that both spectral steepening at ion-kinetic scales and the dissipation mechanisms depend on the nature of energy transfer in the inertial range \cite{Smith2006,Bruno2014}.

The non-universal nature of turbulent dissipation likely has important implications for the acceleration and global structure of the solar wind. In particular, the observation that imbalanced large-scale turbulence dissipates into ion heat via ion-cyclotron fluctuations suggests that turbulence in fast-wind streams, which are known to be more imbalanced, should predominantly heat ions. Conversely, our observation of stronger, more intermittent sub-ion turbulence for less imbalanced intervals suggests that turbulence in lower-Alfv\'enicity slow-wind streams should predominantly heat electrons. Interestingly, because of the electrons' higher thermal velocities, a quantity of energy deposited into ion heat at distances far from the Sun is especially efficient at accelerating the plasma to high speeds, as opposed to simply heating it up \cite{Hansteen1995}. Thus the large-scale turbulence imbalance, by controlling the plasma’s dissipative mechanisms, could directly influence the acceleration efficiency of a parcel of plasma. Such a mechanism could potentially combine with other well-known processes that link turbulence properties and acceleration \cite{Cranmer2009a} to
control the solar-wind’s acceleration and properties in different regions.


These observations have significant ramifications for understanding energy transfer in diverse astrophysical systems such as the intracluster medium \cite{Zhuravleva2014} and accretion disks \cite{Quataert1998}. These systems highly important to our understanding of the universe but can only be observed via radiative signatures. While studies of thermodynamics and particle heating has largely been considered as functions of $T_p/T_e$ and $\beta$ \cite{Quataert1998,Verscharen2022}, our results highlight the importance in understanding the turbulent properties, e.g. the cross helicity, that may affect dissipation and heating processes in these environments. Further work connecting these in situ observations of heating in heliospheric plasmas with astrophysical environments is likely to benefit our understanding of processes occurring broadly in the universe. 


\section{Acknowledgements}
TAB is supported by NASA PSP-GI Grant No. 80NSSC21K1771 as well as the PSP contract NNN06AA01C. CHKC is supported by UKRI Future Leaders Fellowship MR/W007657/1 and STFC Consolidated Grants ST/T00018X/1 and ST/X000974/1.
Support for J.S. was provided by Rutherford Discovery Fellowship RDF-U001804, which is managed through the Royal Society Te Ap\=arangi.

\section{Methods}

We investigate Parker Solar Probe (PSP) observations \cite{Fox2016}for the full day of 2021-09-27. Measurements of the local plasma conditions are made with the PSP Solar Wind Electron Alphas and Proton (SWEAP) experiment \cite{Kasper2016}. We use merged search-coil (SCM) and fluxgate magnetometer data from the PSP FIELDS instrument suite \cite{Bale2016,Bowen2020b}  enabling observation of the turbulent cascade from the inertial to kinetic ranges \cite{Bowen2020c}. Wave-polarization is studied with the fluxgate magnetometers, as the search-coil and merged data only have two functioning components \cite{DudokdeWit2022}. The total electron density is obtained from the FIELDS quasithermal noise measurements \cite{Moncuquet2020}. Data are separated into 747 intervals of approximately 224 seconds (131072 samples at $\sim586$ samples/sec). Each individual interval overlaps 50\% with neighboring intervals to increase the size of the statistical ensemble. The turbulent spectra is determined through the trace power-spectral density (PSD) $E(f)$, of the magnetic field through Fourier methods. 

The Taylor hypothesis is used to convert spacecraft frequency $f$ to wave-number $k$ as $f=2\pi k v_{sw}$, as the solar wind speed is larger than the Alfvén speed. Each spectrum is interpolated onto 56 logarithmically spaced frequencies ranging from $\sim0.06 f_{di}$ to $\sim 12 f_{di}$, where $f_{di}$ corresponds to $kd_i=1$. The maximum scale 12.4$kd_i$ was chosen to avoid the instrumental noise floor at higher frequencies \cite{Bowen2020b}. Below approximately $0.03kd_i$ an artifact is introduced due to the logarithmic interpolation of the PSD. We do not include these scales in our analysis and the artifact does not affect our results. 
The variation in $\alpha$ with wave-number is studied through measuring the local slope of the logarithmically spaced PSD in a 13 point window centered at each frequency. 

A 36-scale Morlet wavelet transform, with response between 0.4 and 18 Hz, is used to identify circularly polarized waves,
\begin{equation} \tilde{B}(s,t)=\sum_{i=0}^{N-1} \psi\left(\frac{t_i-\tau}{s}\right)B(t_i).\end{equation} and The complex-valued wavelet transform is rotated into a field aligned coordinate (FAC) system parallel to the local background field $\mathbf{\tilde{B}}_{FAC}=(\tilde{B}_{\perp1},\tilde{B}_{\perp2},\tilde{B}_\parallel)$. We study signatures of circular polarization using
\begin{align}
    \sigma_B(f,t)=-2\text{Im}(\tilde{B}_{\perp1}\tilde{B}_{\perp2}^*)/(\tilde{B}_{\perp1}^2+\tilde{B}_{\perp2}^2),
\end{align}
 with left/right handed waves corresponding to positive/negative helicity. Waves are identified when $\lvert\sigma_B\rvert > 0.9$ \cite{Bowen2020a,Bowen2020c}. 
 Negative values of $\sigma_B$ correspond to a left-handed rotation of the field, while positive values of $\sigma_B$ correspond to a right handed polarization. 
 The fraction of circularly polarized power is determined by filtering wavelet coefficients with $\lvert\sigma_B\rvert > 0.9$, and normalizing the polarized power $E_{\lvert\sigma\rvert>0.9}(f)$ to the total observed power $E_\perp(f)=\lvert\tilde{B}_{\perp1}^2+\tilde{B}_{\perp2}^2\rvert$.

When the angle between the solar wind flow and the mean magnetic field, $\theta_{BV}$, is sufficiently oblique, observation of quasi-parallel waves is complicated, as the polarization plane of parallel-propagating circularly polarized waves is not aligned with the flow over the spacecraft \cite{Bowen2020a}. Additionally, solar wind turbulence is anisotropic \cite{Horbury2008}, such that observed perpendicular turbulent fluctuations can dominate over parallel-propagating ICWs at oblique $\theta_{BV}$. To control for these observational effects, we consider only intervals with  $15^\circ<\langle\theta_{BV}\rangle<25^\circ$, controlling for anisotropy \cite{Horbury2008}, while capturing a sufficient ensemble of intervals with and without wave signatures. Our results are robust to varying the range of $\theta_{BV}$, even when a large range, e.g. $0^\circ<\theta_{BV}<40^\circ$ is studied. However, the anisotropy associated with large range of $\theta_{BV}$ may affect the results \cite{Bowen2020a}. While single point measurements pose inherent limitations, selecting only intervals $15^\circ<\theta_{BV}<25^\circ$ enables us to observe the scaling properties of the perpendicular turbulent cascade alongside quasi-parallel waves. Although the lack of ability to study the full 3D turbulent cascade in the presence of ion-scale waves is unfortunate, our observations of the quasi-parallel spectra provides significant evidence for the mediation of turbulent dissipation through cyclotron resonance.

In order to relate the presence of waves to observed signatures of  kinetic-scale turbulence, we compute magnetic-field increments,

\begin{equation}
    \Delta B_i^\tau= B_i(t) - B_i(t +\tau),
\end{equation}
 using $\tau$ ranging from $\sim$0.005-7 s ($\sim0.1 -100$ Hz). The probability distribution of increments at a given $\tau$, noted as $P({\Delta B}_i^\tau)$, characterizes signatures in the fluctuations that relate to turbulent dissipation, e.g. non-Gaussianity and intermittency \cite{Sorriso-Valvo1999,Mallet2019}.

Structure functions of the increments of order $p$ are constructed as
\begin{align}
S_i^p (\tau)= \langle \lvert\Delta B_i^\tau\rvert^p\rangle,
\end{align} 
where  $\langle..\rangle$ indicates an average over each 224 second interval. Moments of $P({\Delta B}_i ^\tau)$ are approximately accurate up to $p< \text{log} N -1 $ \cite{DudokdeWit2004}; with $N$=131072,
moments with $p \le 4$ are likely well measured in each interval. Signatures of turbulent heating and dissipation are known to correlate with the non-Gaussianity of the fluctuations \cite{Sorriso-Valvo1999, parashar2015}. Deviations from Gaussianity are often measured using the kurtosis of the fluctuations $\kappa(\tau)_{\Delta B_i} = S^4_i(\tau)/S^2_i(\tau)^2$. A Gaussian distribution has $\kappa=3$, with larger values corresponding to heavy-tailed distributions indicating non-Gaussian structure\cite{Sorriso-Valvo1999, parashar2015}. Intermittency in turbulence is usually defined as increasing non-Gaussianity in $P({\Delta B}_i ^\tau)$ as the scale size decreases \cite{Frisch1995}.

Intermittency is also evident in the scaling of moments of the $P({\Delta B}_i ^\tau)$ distribution over various $\tau$ \cite{Dudokdewit1996,Sorriso-Valvo1999,Kiyani2009,Chhiber2021,Roberts2022}. Non-intermittent scalings in $P({\Delta B}_i^\tau)$ correspond to self-similarity at each $\tau$ with 

\begin{equation}P(\Delta B_i^\tau) = \tau^{-H}f(\Delta B_i^\tau/\tau^H)
\label{eq:selfsim}.\end{equation} 
where $f(x)$ is an arbitrary function and $H$ is some number. The prefactor of $\tau^{-H}$ allows for normalization
\begin{equation}
1 = \int P(\Delta B_i^\tau)d\Delta B \end{equation} at each $\tau$.
Self-similar scaling of $P({\Delta B}_i^\tau)$ results in structure functions that scale as $S_i^p(\tau) \propto \tau^{\zeta(p)}$ with  $\zeta(p)=Hp$  \cite{Sorriso-Valvo1999,Kiyani2009}. In general, if  $P({\Delta B}_i^\tau)$ is not self-similar according to Eq. \ref{eq:selfsim}, then $S_i^p(\tau)\propto \tau^{\zeta(p)}$, where the increasing non-Gaussian structure in intermittent fluctuations results in a nonlinear $\zeta(p)$ \cite{Sorriso-Valvo1999}.  We test for intermittency in kinetic-scale turbulence by measuring $\zeta(p)$ between 20 Hz $<f <$ 100 Hz, corresponding to sub-ion scales. A constant $d{\zeta}/d{p}$ indicates that the kinetic-scale fluctuations show self-similarity whereas a variable $d{\zeta}/d{p}$ indicates intermittency.

We compute the curvature in $\zeta$ for each interval \begin{align}
\Gamma_{\zeta}=\frac{\rvert\zeta''(p)\lvert}{(1+\zeta'(p)^2)^{3/2}},
\end{align}
to attain a single number in the the nonlinearity of the scaling exponents to correlate against other observable parameters. In each interval we measure and report the average $\Gamma$ over $p$. The study was repeated using maximum $\Gamma_{\zeta}$ in each interval, though no singificant variation in the results could be obtained.

Errors on each of these quantities are approximated via bootstrapping methods. We generate a list of 10000 random indices from each N=131072 sample interval. From this sub-sampled interval, we estimate the kurtosis, $\zeta(p)$, and ${\Gamma_{\zeta}}(p)$ using the above methods. This process is repeated 16 times for each interval to give an ensemble of estimated values for $\kappa$, $\zeta(p)$ and $\Gamma_{\zeta}(p)$. The error on the mean estimates are reported as the standard error of this distribution.  For errors on $\zeta(p)$, we do not consider the error on each $p$ for interval individually, but consider both the average and maximum measured error for each $p$, shown in Fig 5(c). We find that error on $\zeta (p)$ increases with $p$, which is consistent with expected behavior due to finite sample effects. Moments up to $p \approx 3$ should be accurately estimated \cite{DudokdeWit2004}.  We find that the average standard error $\langle\epsilon_\zeta\rangle$ at $p=3$ was  0.008 and the maximum measured standard error was $\text{max}[\epsilon_\zeta]=0.05$.  At $p=6$ the average standard error $\langle\epsilon_\zeta\rangle$ was 0.06 and the maximum measured standard error was $\text{max}[\epsilon_\zeta]=0.36$. These errors are relatively small with $\epsilon_\zeta/\zeta \lt \sim 10\%$. This same method is used to determine the standard error on the derivative $d\zeta/dp$, $\epsilon_\zeta'$, which is of order 1\%. Estimation of the error on the curvature $\Gamma_{\zeta}$ is similarly performed using the standard error of the mean value determined from the standard deviation of the 16 element ensemble.

\section{Data Availability Statement}
PSP data are publicly available at NASA Space Physics Data Facility (SPDF) \url{https://cdaweb.gsfc.nasa.gov/}.
FIELDS data are also hosted at \url{https:sprg.ssl.berkeley.edu/data/psp/data/sci/fields/}.
\bibliography{sn-bibliography}

\end{document}